\documentclass[twocolumn, switch]{article} % Method A for two-column formatting

\usepackage{preprint}

\usepackage{amsmath, amsthm, amssymb, amsfonts}

\usepackage[numbers,square]{natbib}
\usepackage[utf8]{inputenc}	% allow utf-8 input
\usepackage[T1]{fontenc}	% use 8-bit T1 fonts
\usepackage{xcolor}		% colors for hyperlinks
\usepackage[colorlinks = true,
            linkcolor = purple,
            urlcolor  = blue,
            citecolor = cyan,
            anchorcolor = black]{hyperref}	% Color links to references, figures, etc.
\usepackage{booktabs} 		% professional-quality tables
\usepackage{nicefrac}		% compact symbols for 1/2, etc.
\usepackage{microtype}		% microtypography
\usepackage{lineno}		% Line numbers
\usepackage{float}			% Allows for figures within multicol
\usepackage{placeins}
\usepackage{tabularx}

\usepackage{newfloat}
\DeclareFloatingEnvironment[name={Supplementary Figure}]{suppfigure}
\usepackage{sidecap}
\sidecaptionvpos{figure}{c}

\usepackage{titlesec}
\titlespacing\section{0pt}{12pt plus 3pt minus 3pt}{1pt plus 1pt minus 1pt}
\titlespacing\subsection{0pt}{10pt plus 3pt minus 3pt}{1pt plus 1pt minus 1pt}
\titlespacing\subsubsection{0pt}{8pt plus 3pt minus 3pt}{1pt plus 1pt minus 1pt}

\usepackage{tikz,xcolor,hyperref}

\definecolor{lime}{HTML}{A6CE39}
\DeclareRobustCommand{\orcidicon}{
	\begin{tikzpicture}
	\draw[lime, fill=lime] (0,0) 
	circle [radius=0.16] 
	node[white] {{\fontfamily{qag}\selectfont \tiny ID}};
	\draw[white, fill=white] (-0.0625,0.095) 
	circle [radius=0.007];
	\end{tikzpicture}
	\hspace{-2mm}
}
\foreach \x in {A, ..., Z}{\expandafter\xdef\csname orcid\x\endcsname{\noexpand\href{https://orcid.org/\csname orcidauthor\x\endcsname}
			{\noexpand\orcidicon}}
}
\title{Deep Reinforcement Learning for Backup Strategies against Adversaries}

\usepackage{xwatermark}
\newwatermark[firstpage,color=gray!90,angle=0,scale=0.28, xpos=0in,ypos=-5in]{ \texttt{*pascal.debus@aisec.fraunhofer.de}}

\usepackage{authblk}

\author[1\thanks{\tt{pascal.debus@aisec.fraunhofer.de}}]{Pascal Debus}
\author[2]{Nicolas M\"uller}
\author[3]{Konstantin B\"ottinger}
\affil[1,2,3]{Fraunhofer AISEC, Garching, Germany}

\begin{document}

\twocolumn[ % Method A for two-column formatting
  \begin{@twocolumnfalse} % Method A for two-column formatting
  
\maketitle

\begin{abstract}
Many defensive measures in cyber security are still dominated by heuristics, catalogs of standard procedures, and best practices. Considering the case of data backup strategies, we aim towards mathematically modeling the underlying threat models and decision problems. 
By formulating backup strategies in the language of stochastic processes, we can translate the challenge of finding optimal defenses into a reinforcement learning problem. This enables us to train autonomous agents that learn to optimally support planning of defense processes. In particular, we tackle the problem of finding an optimal backup scheme in the following adversarial setting:
Given $k$ backup devices, the goal is to defend against an attacker who can infect data at one time but chooses to destroy or encrypt it at a later time, potentially also corrupting multiple backups made in between. In this setting, the usual \emph{round-robin} scheme, which always replaces the oldest backup, is no longer optimal with respect to avoidable exposure. 
Thus, to find a defense strategy, we model the problem as a hybrid discrete-continuous action space Markov decision process and subsequently solve it using deep deterministic policy gradients. 
We show that the proposed algorithm can find storage device update schemes which match or exceed existing schemes with respect to various exposure metrics.
\end{abstract}

%\keywords{First keyword \and Second keyword \and More} % (optional)
\vspace{0.35cm}

  \end{@twocolumnfalse} % Method A for two-column formatting
] % Method A for two-column formatting

%\begin{multicols}{2} % Method B for two-column formatting (doesn't play well with line numbers), comment out if using method A

%%%%%%%%%%%%%%%  Main text   %%%%%%%%%%%%%%%
% \linenumbers

% keywords can be removed
%\keywords{First keyword \and Second keyword \and More}

\section{Introduction}\label{introduction}
An essential component of every IT infrastructure is a reliable backup system which minimizes the damage caused by both natural accidents (fire, disk crash, water) as well as cyber attacks.
While such natural accidents are immediately apparent, cyber attacks aim at stealthy infection and often come in two phases. After infection, the objective (e.g. encryption or deletion of files) is delayed in order to infect newly made backups as well. A well-known example of this is ransom-ware, which encrypts a user's hard drive and tries to extort payment in Bitcoin for the restoration of the files~\cite{bitcoin_paper}.

Thus, a user is faced with the following problem: There is a fixed number of available backup devices, each of which can hold only a single backup of the database at a specific point in time. The user must consecutively decide when and also which of the older backups is overwritten with the most recent state of the database while not knowing if and when an infection has occurred. 

Since backups made after time of infection are corrupted, a good backup strategy has to deal with the conflicting goals of keeping backups available which are firstly old enough to not be infected, while secondly as recent as possible such that not too much information is needlessly lost. In this setting, the so-called round-robin scheme, which always replaces the oldest backup, is no longer optimal with respect to avoidable exposure.

While a single user may easily add more backup devices, this problem is much more relevant for companies operating on large volumes of business critical data. For these, adding more backup devices comes at a significant cost due to the sheer volume of their data, as well as maintenance and privacy concerns. 
Even though the costs of transferring data into long-term cloud storage such as Amazon Glacier are low, this is not a feasible remedy since most of the costs actually occur when data needs to be retrieved again. Depending on the pricing model the retrieval process is usually bandwidth-limited and generally much slower when compared to on-site backup devices. This is a problem when the time to recovery is critical, e.g. in production sites where every hour of outage causes a significant loss.
%such that the direct cost of the data loss is just replaced by the retrieval cost of the long-term backup.
Thus, finding a strategy to optimally utilize a given number of backup devices is of great interest.
By measuring the performance of such a strategy using loss exposure metrics, we can cast the problem as optimization problem and solve it to obtain concrete schemes, as shown by \cite{checkpoint_worst_case}, \cite{bs}. 

Additional value can be added by treating the optimization as learning process of an autonomous agent driven by the loss exposure feedback: Such an agent will not only provide
a performant scheme but will also be able to deal with deviations from that scheme, as they will likely occur in any real world application.

We contribute to solving this problem as follows:
\begin{itemize}
    \itemsep0em
    \item We formalize the problem as a hybrid discrete-continuous action space Markov decision process and thus transform it to a reinforcement learning problem.
    \item We propose an algorithm that learns to find update sequences and time steps which minimize mean avoidable exposure metrics based on an extension of deep deterministic policy gradients for hybrid actions spaces.
    \item We demonstrate that the algorithm produces backup sequences and steps sizes that perform better with respect to mean avoidable exposure metrics than the reference algorithm that strictly enforces a worst-case optimal lower bound at \emph{every} step.
    \item Finally, to stipulate further research, we publish our implementation.\footnote{Repository link available upon publication}
\end{itemize}

The paper is structured as follows: Section~\ref{sec:relatedWork} summarizes related work on checkpointing algorithms and deep reinforcement learning. Section~\ref{DRL} describes the relevant theory of deep reinforcement learning algorithms,
in particular deep deterministic policy gradients (DDPG). Subsequently, Section~\ref{mdp_model} presents the formalization of the backup problem and how to translate it to a Markov decision process. In Section~\ref{sec:implementation} we present details on the implementation and necessary modifications of DDPG to apply it to hybrid discrete-continuous actions spaces. Section~\ref{results} shows the results of applying the suggested algorithm to find schemes for some concrete $k$ and discusses the results. Finally, Section~\ref{conclusion} concludes the paper and suggests some further research directions.

\section{Related Work}\label{sec:relatedWork}
\subsection{Checkpointing Algorithms and Backup Strategies}
The research on checkpointing and backup strategies is a very large and diverse field with varying objectives. Some early work like \cite{young_first_1974,gelenbe_optimum_1979} focuses on finding the optimal interval length between checkpoints/backups based on some stochastic model of fault distribution or system load in order to maximize availability. \cite{nakamura_optimal_2003} considers backup schemes based on incremental and full backups and optimizes the length between full backups given a cost model for the overhead of incremental and full backups. Other approaches such as \cite{lim_power-aware_2011} concentrate on very specific environments like remote backups of mobile devices and try to optimize power consumption. 
In summary, most related work is centered around finding \emph{the} single optimal interval length in an equidistant setting with respect to some metric and environment model. In contrast, there is not much work on strategies in an adversarial setting and the more realistic case of a limited number of non-equidistant checkpoints.

The problem of finding an optimal backup strategy in the face of cyber attacks has been first brought up by \cite{bs} in the more general context of online checkpointing algorithms and was initially mathematically formalized in \cite{checkpoint_worst_case}.  A checkpoint is any memorized state of a computation system that can be used to roll back computation from time $T$ to an earlier time, $T_{\text{checkpoint}} < T$, without having to restart from the system's initial state. 
Other applications are, for example, the execution state of a program in an interactive debugging session: While looking for the root cause of the bug, the developer might accidentally step over the problematic code sections and is then faced with the problem of reproducing the execution state just before the problem. In this setting, avoidable exposure does not refer to a loss of data but the loss of working time when having to restore execution states from a very old checkpoint.

The problem of minimizing avoidable exposure has its roots in mathematical discrepancy theory (see, for example, \cite{beck1987}) which tries to characterize unavoidable deviations of sequences from a uniform distribution. A natural metric to measure discrepancy is the worst case ratio between actual and ideal interval lengths formed by the sequence. 

In their work, Bar-On et al.~\cite{bs} identify necessary and sufficient properties that must hold for schemes with \emph{worst-case discrepancy guarantees} and subsequently derive tight upper and lower discrepancy bounds for checkpointing schemes on $k\leq 10$ checkpoints as well as an asymptotic upper and lower discrepancy bound of $\ln 4$ for large $k$. 
These bounds are also computationally verified and an algorithm is provided which generates candidate periodic update sequences. The corresponding step size sequence is then computed by finding feasible solutions to a finite subprogram of an infinite linear program whose constraint set is given by the initially proven properties.

In contrast to this, our approach uses reinforcement learning to learn a policy that minimizes the \emph{mean discrepancy} for an arbitrary initial or intermediate set of checkpoints.

\subsection{Deep Reinforcement Learning}\label{subsec:relatedWork_drl}
We solve the backup problem using an extension of the deep deterministic policy gradient (DDPG) algorithm. This algorithm was initially proposed by \cite{ccwdr} and was used to simulate a number of physical tasks, such as cartpole swing-up and car driving. DDPG itself is an extension of the \emph{Deep Q Network} (DQN) algorithm by \cite{mnih}, which has been widely popularized due to its human level performance in Atari Video Games. 

Deep Q-Learning made it possible to effectively handle continuous or very large discrete state spaces using neural network based value-function approximation. DDPG extended this approach to continuous action spaces by using a policy network that is connected to the DQN within an actor-critic infrastructure\cite{sutton_ac}. 

Finally, Hausknecht and Stone \cite{param_act_space} suggested a method to deal with parameterized action spaces, i.e. a set of discrete actions augmented by one or more continuous parameters, which they applied to the domain of simulated RoboCup soccer. In this approach, the discrete action with the highest score is chosen, along with its parameters. 

For our implementation, we also follow this general idea but treat the step size as independent output instead of a parameter of the discrete action such that the action space can be considered hybrid rather than parameterized.

\section{Background: Deep Reinforcement Learning}\label{DRL}
In this section, we provide the necessary background on reinforcement learning.
The framework of Markov decisions processes (MDPs) is usually the best way to formally analyze most reinforcement learning algorithms. A Markov decision problem is defined
as the tuple $(\mathcal{S}, \mathcal{A}, P, r, \gamma)$, where $\mathcal{S}$, $\mathcal{A}$ denote the state space and action space, respectively, $P$ the state transition probability matrix, $r$ the reward function and lastly, $\gamma$ is the discount factor for future rewards.

\subsection{Q-Learning and Bellman equation}
Consider an MDP with state space $\mathcal{S}$ and action space
$\mathcal{A}=\mathbb{R}^n$. Let $E$ denote the (potentially stochastic) environment which determines state distribution, transition as well as reward dynamics.

The agent's behavior is determined by a (generally stochastic) policy $\pi$, which assigns to each state $s\in\mathcal{S}$ a probability distribution $\mathcal{P}(\mathcal{A})$ over all possible actions.
At the core of many reinforcement learning algorithms are value functions such as the action-value function $Q^{\pi}(s_t,a_t)$ which is defined in equation~(\ref{eq:bellman_stoch}) to be the expected total discounted reward $R_t$ for each state-action combination at time $t$ when following the policy $\pi$ after the first action $a_t$. 
\begin{align}
    \label{eq:bellman_stoch}
    Q^{\pi}(s_t,a_t) &= \mathbb{E}_{\substack{r_{i\geq t}, s_{i>t}\sim E\\ a_{i>t}\sim\pi}}[R_t|s_t, a_t] \\
    &= \mathbb{E}_{r_t,s_{t+1}\sim E}[r(s_t,a_t) + \gamma \mathbb{E}_{a_{t+1}\sim E}[Q^{\pi}(s_{t+1},a_{t+1})]] \nonumber
\end{align}

The second equality in equation~(\ref{eq:bellman_stoch}) is also known as Bellman equation and provides a way to compute the expected reward values recursively. 
In the case of a deterministic policy $\mu: \mathcal{S}\to\mathcal{A}$, the action is known to be $\mu(s_{t+1})$ and the Bellman equation can be simplified to 
\begin{equation}
    \label{eq:bellman_determ}
    Q^{\mu}(s_t,a_t) = \mathbb{E}_{r_t,s_{t+1}\sim E}[r(s_t,a_t) + \gamma Q^{\mu}(s_{t+1}, \mu(s_{t+1}))].
\end{equation}
Since the action is not a random variable any more, the expectation does not depend on it which enables the agent to learn the value function for $\mu$ \emph{off-policy}, i.e. by following a stochastic exploration policy. 
If $\mu$ is the \emph{greedy policy}, $\mu(s)=\arg\max_a Q(s,a)$, this algorithm is known as Q-learning. For large state spaces one usually has to resort to parameterized approximations $Q^{\mu}(s,a|\theta^Q)$ of the value function, using a (deep) neural network with weights $\theta^Q$, which is consequently referred to as deep Q-learning.

Following the greedy policy for continuous $a$, however, poses a problem because finding the $\arg\max$ is essentially an optimization problem that needs to be solved for every step.

\subsection{Deep Deterministic Policy Gradient (DDPG)}
The DDPG algorithm avoids this problem by also using a parameterized approximation $\mu(s|\theta^\mu)$ of the deterministic policy which is connected to the action-value function approximator by the actor-critic architecture \cite{sutton_ac}. The critic learns the action-value function using deep Q-learning, while the actor uses gradient ascent to learn weights $\theta^\mu$ such that the performance objective $J(\mu^{\theta})=\mathbb{E}_{s_i\sim \rho^\mu}[R_1]$ is maximized, where $\rho^\mu$ denotes the density of the discounted state distribution (when following $\mu$).

The quantitative relation between the actor and the critic approximators and their gradients is given by the \emph{Deterministic Policy Gradient Theorem} stated by \cite{dpg_silver}:
\begin{equation}
\label{eq:on_policy_dpgt}
    \nabla_{\theta^\mu}J(\mu^{\theta}) = \mathbb{E}_{s_t\sim\rho^\mu}[\nabla_{\theta^\mu}Q^{\mu}(s,a|\theta^Q)|_{s=s_t,a=\mu(s_t|\theta^\mu)}]
\end{equation}
For an \emph{off-policy actor-critic method}, however, which uses a stochastic exploration policy $\pi$, this relation only holds approximately:
\begin{equation}
\label{eq:pff_policy_dpgt}
    \nabla_{\theta^\mu}J(\pi) \approx \mathbb{E}_{s_t\sim\rho^\mu}[\nabla_{\theta^\mu}Q^{\mu}(s,a|\theta^Q)|_{s=s_t,a=\mu(s_t|\theta^\mu)}]
\end{equation}

\FloatBarrier
\section{Backup Strategy as Markov Decision process}\label{mdp_model}
In this section we show how the problem can be formulated as Markov decision process (MDP). In the following we first summarize the original formalization and important concepts introduced by \cite{checkpoint_worst_case} and \cite{bs}, then transform them to the domain of MDPs by defining scale invariant state and action spaces, the transition dynamics and the reward function .

\subsection{Formalization of the backup problem}\label{optimal_backup}
A backup scheme based on $k$ storage devices is fully determined by an infinite sequence of update actions $(d_n, t_n)$, where $t_n$ denotes the time when the $n^{\text{th}}$ backup is written on the storage device specified by $d_n$. Hence, $t_n$ is an unbounded, monotonically increasing sequence while $d_n$ is a label from $\{0,1,\ldots,k\}$. Since the storage devices are assumed to be identical and differ only by the age of the backup, they can be relabelled after each update action such that the label $l\in\{0,1,\ldots,k\}$ always refers to the $(l+1)^{\text{th}}$ oldest backup.
\subsubsection{Snapshots and Updates}
\begin{figure}[h]
    \centerline{\includegraphics[width=\columnwidth]{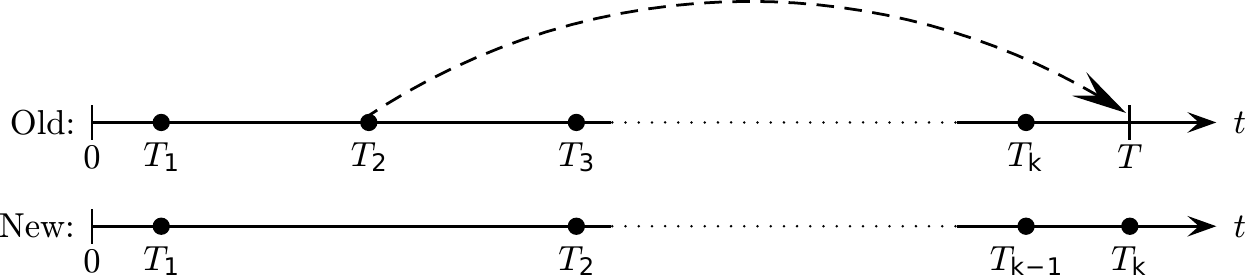}}
    \caption{Illustration of update action $(d,t)=(2,T)$. Source: \cite{bs}}
    \label{fig:update_action}
\end{figure}

At any time $T$, the state of the backup scheme can be be described by a snapshot $S = (T_1, T_2, ..., T_k)$ of times at which each backup device was last updated. Consequently, an update action $(d,t)=(2,T)$ will remove $T_2$ from the snapshot, decrement the label of all devices with label $l>2$ and append $T$ as new $T_k$ to the snapshot, as illustrated by Figure \ref{fig:update_action}.

\subsubsection{Attack Scenario}
\begin{figure}[h]
    \centerline{\includegraphics[width=\columnwidth]{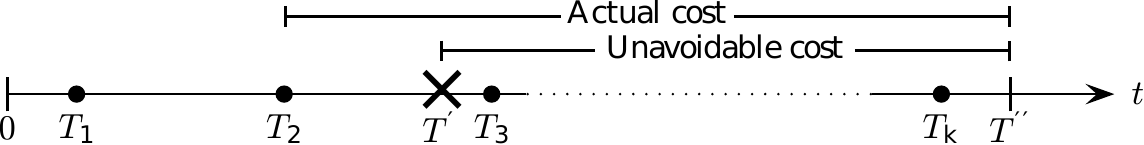}}
    \caption{Unavoidable cost for attack with $T'\in [T_2,T_3]$. Source: \cite{bs}}
    \label{fig:unavoidable_cost}
\end{figure}

In an attack scenario, there are two distinct time points: a secret infection time $T'$
and an execution time $T''$ that will become immediately apparent. The total recovery cost is then proportional to $T''-T_i$ where $i=\arg\max_k\{T_k|T_k < T'\}$, which corresponds to resetting the system to the backup fabricated at time $T_i$. The part of the cost proportional to  $T'' - T'$ is unavoidable, since backups fabricated during $[T', T'']$ are corrupted as well such that only the part proportional to $T' - T_i$, the avoidable loss, can be subject to minimization, as shown in Figure \ref{fig:unavoidable_cost}.

\subsubsection{Threat Model}\label{threat_model}
It is the attacker's objective to maximize the defender's recovery cost, either to increase the ransom that can be demanded or to increase the damage to the defender's assets. Due to the information asymmetry, additional cost can be inflicted on the defender by having the infection
time $T'$ \emph{just before} the next scheduled update $T_{k+1}$ is performed. Moreover, this extra cost is maximal if $[T_{k}, T_{k+1}]$ happens to be the largest interval in the snapshot once $T''$ is reached.

Hence, the achievable additional cost depends on the attacker's knowledge of the backup schedule. For an attacker from the outside it is reasonable to assume that $T'$ is uniformly distributed within every given interval $[T_{i}, T_{i+1}]$ due to lack of knowledge of the schedule, uncertainties through the infection vector, or both. For an insider, on the other hand, the distribution will be biased towards $T_{i+1}$ and will be most likely concentrated immediately before $T_{i+1}$ for an attacker with perfect knowledge.

In any case, the defender seeks to minimize this additional exposure but has to trade off strict worst case exposure guarantees for lower average exposure depending on the expected attacker knowledge. Intuitively, this means that for attackers with limited knowledge, the defender will achieve a lower loss on average than if he had committed to the bounded worst-cast scheme. In the case of a perfect knowledge attacker the loss will of course be lower for the latter scheme.

\subsubsection{Discrepancy and $q$-efficiency}
From the perspective of the defender, the worst case avoidable loss is realized if an attack occurs with $T'$ located right before the end of the \emph{longest} interval of any given snapshot $S$. Since the longest interval is minimal for snapshots with exactly uniformly distributed update times, the optimal strategy for the defender is to keep the snapshots uniform at all time. However, due to the fact that only one backup can be replaced per update action, this is not possible for $k>2$: Starting from an initially uniformly distributed snapshot, any update action will result in a snapshot with an interval at least twice as long as all others.

Thus, deviations from the uniform distribution are unavoidable which links the problem to discrepancy theory and leads to the important concepts of $q$-\emph{compliance} and $q$-\emph{efficiency}. An interval defined by update times $T_i-T_{i-1}$ is $q$-compliant at time $T$ if $T_i-T_{i-1}\leq q\frac{T}{k}$. As a consequence, the $q$-value can be interpreted as the ratio between actual and ideal interval length. Finally, a backup scheme is $q$-efficient if all intervals in all its snapshots are $q$-compliant. It is clear that this natural discrepancy metric $q$ needs to be minimized to obtain a good backup scheme.

Finally, it must be discussed if assuming the presented \emph{security} threat model and minimizing exposure with respect to the just introduced metric will have adverse effects on \emph{safety} threats such as naturally occurring data loss due to disk crash, fire or similar natural hazards. However, in this case we just have $T^{'}=T^{''}$ and the exposure reduces to the length of just the \emph{last} interval. Since the schemes try to keep the snapshots as uniform as possible, this will naturally hold also for the last interval: In half of the cases the last interval might a bit longer than the other intervals in the snapshot and in the half a bit shorter such that on average the scheme is neutral with respect to safety threats.

\subsection{State Space}
The concept of the snapshot translates almost directly to the concept of a state in a Markov decision process (MDP). However, some adjustments are necessary since the times specified in the snapshot are unbounded which leads to conceptual problems in the design of the reward function as well as numerical problems concerning the implementation.

With $k$ given backup devices and the goal of keeping the backups as uniformly distributed as
possible the time periods between new updates grow basically at the same rate as the periods between backup times that are part of the snapshot. Therefore, we consider the problem scale-invariant which motivates the following state definition.
\begin{equation}
\label{eq:state_space}
\mathcal{S}:= \left\lbrace\left(S_1, S_2, \ldots, S_k\right) \in (0,1)^k\;\bigg|\; \sum_{i=1}^k S_i=1 \right\rbrace.
\end{equation}
With this state representation, $S_i=T_i-T_{i-1}$ (with $T_0:=0$) represents the time period between two consecutive backups, normalized to the unit interval.
After each update action, the vector is renormalized such that its components sum to one again. This avoids endlessly growing numbers and also ensures that the state-based reward calculation always produces rewards on the same scale.

\subsection{Action Space}
The available actions also follow directly from the problem statement and just need to undergo a consistent scaling scheme. The action space is the product of a discrete and a continuous space: 
\begin{align}
\label{eq:action_space}
\mathcal{A}&:= \mathcal{A}_d\times \mathcal{A}_c = \left\lbrace a=(d, D)\in \{0,1,\ldots,k-1\}\times (1,2] \right\rbrace
\end{align}
For every step one has to to decide \emph{which} (discrete action) old backup will be overwritten and \emph{when} (continuous action) it will be overwritten. We restrict the discrete choice to the devices with label $d\in\{0,1,\ldots,k-1\}$, explicitly excluding the most recent backup devices because, as also shown by \cite{bs}, no $q$-efficient scheme will update it, which is also intuitively clear.

The continuous action parameter $D$ is restricted to the domain $(1,2]$ and can be interpreted as the rescaling factor for the next state. This restriction is not very strict, because choosing a next step of $2$ results in a state where the last interval in the snapshot is as big as the union of all remaining intervals, consequently leading to a very non-uniformly distributed state in the next step.

\subsection{Transitions}
Finally, the transition to a new state is performed by merging the $d^{\text{th}}$ interval of $S$ with its successor, appending the new time step $D$ at the end and subsequently dividing all other components by $D$:
\begin{align}
\label{eq:state_transition}
P_{SS'}^a&: \mathcal{S}\times\mathcal{A}\to \mathcal{S} \nonumber\\
(S,a)&\mapsto P(S,a) =  S' = (S'_1, S'_2, \ldots, S'_k)\\
S' &= \left(\frac{S_1}{D}, \ldots, \frac{S_d+S_{d+1}}{D},\frac{S_{d+2}}{D}, \ldots, \frac{S_k}{D}, (D-1)\right). \nonumber
\end{align}
The transition dynamics defined this way are completely deterministic.

\subsection{Reward Functions}
% There are multiple choices for a reward function given the problem stated in Section \ref{optimal_backup}.
% Since the worst-case loss given perfect attacker knowledge is equal to the largest interval between to consecutive backups in the snapshot, an obvious choice is to use pick the negative length $r_{\text{IL}}:= -\max_{i}s^{(i)}$ of this interval as reward. However, when comparing cumulative rewards of the schemes in \ref{table:qefficient_results} and the round robin scheme, the latter performs actually better.
Since the $q$-value introduced in the notion of $q$-compliant snapshots measures the deviation from equidistribution (i.e. the discrepancy) of the update time sequence, we will base the reward function on the inverse of the $q$ value in order to encourage low discrepancy throughout the updates:
\begin{align}
    \tilde{r}: \mathcal{S} \to \mathbb{R}, \quad
    \tilde{r}(S) = \frac{1}{k\max_i S_i} \in \bigg(\frac{1}{k}, 1\bigg]
\end{align}
To improve stability of learning and make rewards for different values comparable, we subsequently scale rewards to
$(-\lambda,\lambda]$ using a hyperparameter $\lambda>0$ such that we arrive at
\begin{align}
\label{eq:reward}
    r(s,\lambda) &= -\lambda + \frac{2\lambda}{k-1} \left(\frac{1}{\max_i S_i}-1\right).
\end{align}

Finally, since optimal policies for an MDP are those which minimize the \emph{expected}
(discounted) reward, there is a subtle difference in the optimization objective compared to \cite{bs}. The algorithm will not strictly stay below a given discrepancy bound, but minimize the \emph{mean} discrepancy of the resulting backup scheme.

\FloatBarrier
\section{Implementation}\label{sec:implementation}
In this section we present details on the implementation of the proposed algorithm. Since the algorithm is based on DDPG, we start by showing how we incorporate the hybrid action space and the resulting implications for discrete exploration this brings about. Finally, we show the overall network structure and the employed stability-improving techniques. An overview of all hyper parameters and their respective values for our experiments can be found in Table~\ref{table:hyper_params}.

\subsection{Discrete Action Relaxation and Exploration}
The DDPG algorithm is designed to output continuous actions. To tackle the problem of
the discrete part of the hybrid action space, we encode the $k-1$ different actions
using one-hot vectors and subsequently relax the problem by allowing continuous
values $a^{(d)}_i\in [0,1]$ in each component of the vector.
The normalized vector can then be interpreted as a probability distribution $\mathcal{P}$ over the discrete actions $a^{(d)}_i \in \{ 0,1,\ldots,k-2\}$ which is easily produced, e.g. by the
actor network using softmax activation for the output layer.

At this point it is possible to choose the next action $d$ stochastically by sampling from the distribution induced by $a^{(d)}$. Nevertheless, since the overall model of the problem is deterministic, we decided to choose $d$ deterministically by $d = \arg\max_{i} a_i^{(d)}.$
To ensure sufficient exploration, on the other hand, we make the discrete policy $\varepsilon$-greedy and sample uniformly from the discrete action space in each exploration step.

\subsection{Continuous exploration}
Concerning exploration in the intrinsically continuous action space, we use the following
stochastic exploration policy, 
\begin{equation}
    \pi(s_t|\theta_t^\mu) := \mu'(s_t|\theta_t^\mu) = \mu(s_t|\theta_t^\mu) + \mathcal{N},
\end{equation}
where the noise distribution $\mathcal{N}$ is given by an Ornstein-Uhlenbeck process with parameters $\theta=0.3$ and $\sigma=0.1$ as suggested by \cite{mnih}.

\subsection{Network Architecture}

\begin{figure}[h]
    \centerline{\includegraphics[width=\columnwidth]{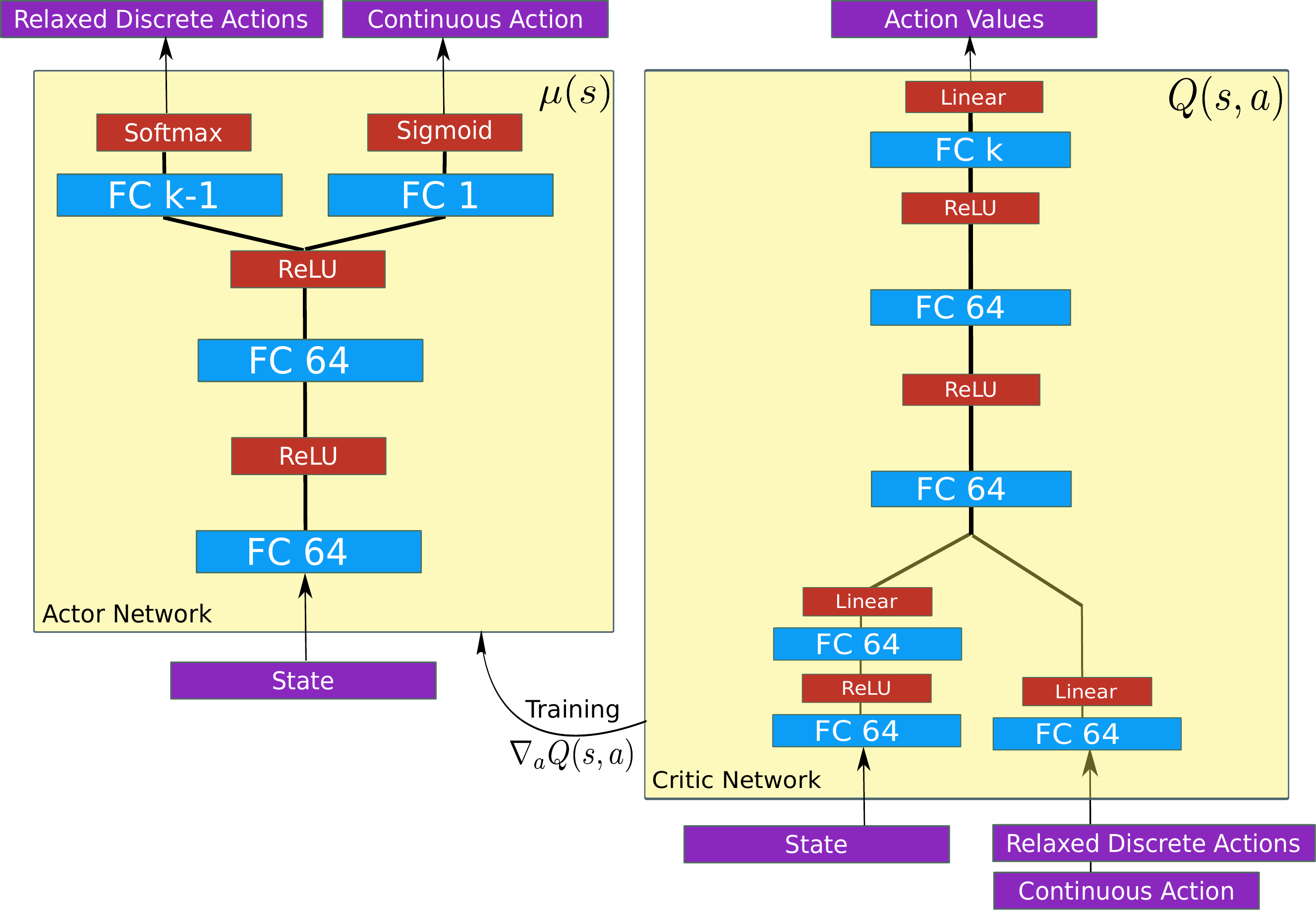}}
    \caption{Architecture of actor and critic networks.\label{fig:network_architecture}}
\end{figure}

The network architecture is shown in Figure \ref{fig:network_architecture} and follows the actor-critic paradigm.
The core of both actor and critic network consists of two fully connected layers with each 64 units using rectified linear activation functions (ReLUs). The output layer of the critic network has $k$ units with linear activations, whereas for the actor network the output uses softmax activation for the first $k-1$ units, representing the relaxed discrete actions, and sigmoid activation for the last unit, representing the continuous step size.

Concerning the input, the $k$-dimensional state vector is passed to the first layer of both networks. For the critic network, however, there are additional fully connected layers with 64 units and ReLU activation and 64 units and linear activation that preprocess the state input, as well as a fully connected layer with 64 units which takes the $k$-dimensional relaxed action representation as input. The output of those two preprocessing layers are subsequently concatenated and passed to the core network described above. 

Since the number of units in the hidden layer is rather small, we also experimented with some variations of this architecture, such as fully connected layers with up 512 units and also tried to improve the training process using priority replay \cite{schaul2015prioritized} or Monte Carlos tree search algorithms \cite{browne_survey_2012}. Nevertheless, none of these modifications lead to a significant improvement of the resulting policy or stability of the training.

% \begin{table}[h!]
% \begin{center}
%     \begin{tabularx}{\textwidth}{lXl|lXl}
%         \toprule
%             &  \textbf{Description}  & \textbf{Value} &   & \textbf{Description}  & \textbf{Value}\\
%         \midrule
%         $N_{\text{eps}}$ & number of episodes & 600 & $\Delta\epsilon_d$ & epsilon increments for linear annealing& 1/(3e6) \\
%         $N_{\text{steps}}$ & number of actions per episode & 10000 & $\epsilon_{c,\text{min}}$ & min. epsilon for continuous exploration & 0.1\\
%         $\lambda$ & reward scale & 5.0 &         $\Delta\epsilon_c$ & epsilon increments for linear annealing& 1/(3e6) \\
%         $l_r$ & learning rate critic & 0.001 & $\mu_{OU}$ & Ornstein-Uhlenbeck (OU) mean & 0.0 \\
%         $l_a$ & learning rate actor & 0.0001 & $\theta_{OU}$ & OU mean reversion speed & 0.3 \\
%         $\tau$ & lag parameter target update & 0.001 & $\sigma_{OU}$ & OU standard deviation & 0.1 \\
%         $\gamma$ & reward discount rate & 0.99 & $N_{B}$ & size of the replay buffer &1000000 \\
%         $\epsilon_{d,\text{min}}$ & min. epsilon $\epsilon$-greedy exploration (discrete) & 0.1 & $N_{b}$ & Minibatch Size &32 \\
%         \bottomrule
%     \end{tabularx}
%         \caption{Hyperparameters\label{table:hyper_params}}
% \end{center}
% \end{table}

\subsection{Stability}
To achieve a stable and convergent learning processes we employ the following techniques.
First, as already shown in equation (\ref{eq:reward}), we transform all rewards to the same scale $(-\lambda,\lambda]$. In our experiments a value of $\lambda=5$ leads to a stable learning process across all $k$, while for larger values the softmax output often got stuck producing very low entropy discrete action distributions resulting in update actions that would always update the same device.

Second, we use target networks for both the actor and the critic networks, where weights are updated towards the original networks' weights with some lag parameter $\tau=0.001$ according to
$\theta^{Q',\mu'} = \tau\theta^{Q,\mu} + (1 - \tau)\theta^{Q',\mu'}$,
% \begin{align}
%  \label{eq:target_networks}
%   \theta^{Q',\mu'} &= \tau\theta^{Q,\mu} + (1 - \tau)\theta^{Q',\mu'},
%  \end{align}
as suggested by \cite{mnih}.
Finally, again following the approach in \cite{mnih}, a replay buffer of size $1000000$ is used, from which mini-batches of size 32 are uniformly sampled to update the networks.

\FloatBarrier
\section{Results}\label{results}

In this section we present the results of applying our algorithm to the backup problem for some values of $k$ and compare them to schemes suggested in \cite{bs}.

\subsection{Evaluation Method}
A straightforward way of evaluating the scheme would be a simulation where we just sample various attack times $T^{'}, T^{''}$ and measure the resulting (avoidable) data loss for multiple runs of the backup scheme. However, this would result in measurements where the maximally possible damage inflicted on the defender is not realized most of the times which makes the algorithms hard to compare. Since we are comparing against a worst-case guarantee baseline algorithm, we hence wish to make the setup as adversarial as possible.

Taking the threat model described in Section \ref{threat_model} into account, this means that a worst case (perfect knowledge) attack (i.e. $T^{'}$ happened just \emph{before} the end of the largest interval in the current snapshot) is assumed to become known (i.e. $T^{''}$ is revealed) after each update action. The most important metrics in this evaluation are the $q$-values which measure discrepancy and therefore the maximal extra cost inflicted on a defender. The $q$-values are computed for every snapshot resulting from an update action and then aggregated to mean $q$-values ($\bar{q}$) and max $q$-values ($\max q$). Consequently, $\max q$ measures the maximal extra cost over the whole run time of the scheme where the attacker can choose the worst interval in the \emph{worst snapshot}. Put differently, when looking at all sets of $k$ consecutive backups, what was the worst additional loss that could have occurred. $\bar{q}$ measures the average extra costs where the attacker can only choose the worst interval in \emph{some snapshot}, i.e. if $k$ consecutive backups are chosen at random, what is additional loss that could have occurred on average.

The evaluation of the algorithms are performed by running the resulting schemes with $250$ update actions and collecting information on the state sequence, update actions (step size and backup device) and the obtained reward in order to compute summary statistics. Since the initial state is always uniformly distributed, which, as discussed, leads to high $q$-values in the following step(s), we exclude the first $50$ actions from the maximum computation.
The validation run results for the proposed algorithm are displayed in Table \ref{table:policy_results}, whereas the corresponding values for the $q$-efficient schemes from \cite{bs} are displayed in Table \ref{table:qefficient_results}. The columns report mean reward $\overline{r}$, mean discrepancy $\overline{q}$, maximum discrepancy $\max q$, mean step size $\overline{D}$ and the sequence of update devices.

\begin{table}[t]
\begin{center}
\begin{tabularx}{\columnwidth}{lcccX}
    \toprule
      $k$     & $\overline{r}$ & $\overline{q}\;(\max q)$ & $\overline{D}$ & Sequence \\
    \midrule
2  &	5.000 &	1.500 (1.500) &	2.000 &	(0) \\
\hline
3  &	3.086 &	1.528 (1.528) &	1.618 &	(0) \\
\hline
4  &	2.477 &	1.540 (1.540) &	1.346 &	(0, 2) \\
\hline
5  &	2.689 &	1.471 (1.471) &	1.325 &	(0, 2) \\
\hline
6  &	2.289 &	1.506 (1.513) &	1.242 &	(0, 1, 2, 0, 2, 4) \\
\hline
7  &	2.316 &	1.484 (1.497) &	1.209 & (0, 2, 3, 0, 4, 2) \\
\hline
8  &	2.259 &	1.478 (1.485) &	1.161 &	(0, 1, 3, 6, 4, 2, 0, 6, 4, 2, 6, 0, 3, 1,  3, 4)\\
\hline
9  &	2.337 &	1.455 (1.475) &	1.163 &	(0, 4, 2, 4, 0, 4, 5, 2)\\
\hline
10 &	2.189 &	1.469 (1.474) &	1.133 &	(0, 4, 2, 4, 0, 4, 5, 2, 0, 4, 8, 2, 4, 8)\\
\hline
11 &	2.183 &	1.462 (1.466) &	1.124 &	 (0, 5, 2, 5, 8, 0, 5, 1, 2, 4, 5, 0,  5, 1, 9, 5, 2, 5)\\
    \bottomrule
    \end{tabularx}
\end{center}
\caption{Results for schemes from \cite{bs} (reference method).\label{table:qefficient_results}}
\end{table}

\begin{table}[t]
\begin{center}
\begin{tabularx}{\columnwidth}{lcccX}
\toprule
    $k$     & $\overline{r}$ & $\overline{q}\;(\max q)$ & $\overline{D}$ & Sequence \\
    \midrule
2  &	5.000 &	1.500 (1.500) &	2.000 &	(0) \\
\hline
3  &	3.084 &	1.528 (1.528) &	1.618 &	(0) \\
\hline
4  &	3.309 &	1.460 (1.664) &	1.415 &	(0, 2) \\
\hline
5  &	2.688 &	1.472 (1.473) &	1.324 &	(0, 2) \\
\hline
6  &	2.968 &	1.442 (1.753) &	1.262 & (0, 2, 2) \\
\hline
7  &	2.522 &	1.456 (1.582) &	1.222 &	(0, 3, 1, 2, 3, 0, 3) \\
\hline
8  &	2.758 &	1.435 (1.766) &	1.188 &	(0, 4, 2, 4) \\
\hline
9  &	2.430 &	1.445 (1.578) &	1.166 &	(0, 2, 3) \\
\hline
10 &	2.692 &	1.432 (1.838) &	1.148 &	(0, 5, 3, 2, 3) \\
\hline
11 &	2.239 &	1.473 (1.656) &	1.139 &	(0, 1, 4, 4, 5) \\
\hline
12 &	2.315 &	1.453 (1.685) &	1.121 &	(0, 2, 5, 6, 0, 6, 1, 6, 2, 6, 2, 6) \\
\hline
13 &	1.114 &	1.704 (1.928) &	1.135 &	(6, 4, 6, 2, 6, 6, 4, 6, 2, 6) \\
\hline
14 &	2.257 &	1.456 (1.981) &	1.091 &	(6, 4, 6, 2, 0, 6, 4, 6, 2, 6, 0, 6, 2, 6, 3, 6, 0, 6, 2, 3, 4, 0, 6, 4, 12)${}^\dagger$ \\
\hline
15 &	1.987 &	1.494 (1.766) &	1.099 &	(0, 7, 4, 7, 0, 4, 7, 1, 7, 4, 7, 1, 7, 4, 7, 0, 7, 4, 7, 0, 4, 7, 0, 7, 4)${}^\dagger$ \\
\hline
16 &	1.959 &	1.497 (1.768) &	1.091 &	(0, 8, 2, 5, 8, 2, 8, 5) \\
\hline
17 &	1.745 &	1.526 (1.754) &	1.087 &	(0, 8, 3, 8, 3, 8, 0, 8, 3, 8) \\
\hline
18 &	1.653 &	1.558 (1.996) &	1.084 &	(8, 2, 8, 8, 0, 8, 2, 8, 2, 8, 0, 8, 2, 8, 2, 8, 8, 2, 8, 0, 8, 2, 8, 2, 8)${}^\dagger$ \\
\hline
19 &	1.414 &	1.611 (1.809) &	1.078 &	(0, 7, 8, 4, 0, 7, 8, 4, 7) \\
\hline
20 &	1.517 &	1.582 (1.992) &	1.074 &	(8, 3, 0, 7, 7, 3, 7, 8, 0, 7, 3, 7, 8, 8, 3, 0, 7, 3, 7, 7, 8, 3, 7, 0, 7)${}^\dagger$ \\
% \hline
% 12 &	2.315 &	1.453 (1.685) &	1.121 &	(0, 2, 5, 6, 0, 6, 1, 6, 2, 6, 2, 6) \\
\bottomrule
    \end{tabularx}
    \end{center}
\caption{Results for the learned policy (proposed method).\label{table:policy_results}}
\end{table}

\subsection{Discussion}
The first thing that becomes immediately apparent is that both algorithm produce almost identical schemes in terms of backup device sequence and step size for $k=2,3,4,5$. In particular, for the case $k=3$ for which it was proven in \cite{bs} that the step size must be equal to the golden ratio $\frac{1+\sqrt{5}}{2}\approx 1.618$, we can also observe convergence to this value throughout the learning. 
The only significant deviation in performance between the algorithms happens for the case $k=4$. In this case, mean reward and consequently mean $q$-value are much better for the learning algorithm, at the cost of a higher $\max q$ value. Hence, this is the first instance, where the trade-off between mean $q$-values and maximum $q$ value can be observed.

Next, for all $k>5$ similar effects can be observed. The proposed algorithm performs consistently better for all the mean metrics at the cost of a higher maximum $q$-value. We can also observe that for these larger values of $k$, our algorithm produces different update device sequences. For most of the produced sequences a periodic pattern can be observed, however others do not show periodicity within a series of 25 consecutive updates. 
The case $k=13$ produces the worst results in terms of mean and maximum discrepancy because it did not converge to a meaningful sequence which must, as also shown by \cite{bs}, contain the oldest backup device with label $0$.

Finally, the aforementioned trade-off between optimizing for worst-case and average case is also present in many other fields and often average-case optimality is preferred: Quicksort and the Simplex algorithm are just two examples of algorithms with great average case performance in their respective fields that are employed very often in practice despite their non-optimality in the worst case. Bringing the discussion back to applications in the security on might ask what knowledge and capabilities attacker most likely has with respect to the threat model discussed in Section~\ref{threat_model}. An almost perfect knowledge attack occurs most likely for the case of an advanced persistent threat or an insider with elevated privileges and in this case the whole backup infrastructure along with all the backups might very well be also within the attacker's reach such that the task of minimizing \emph{avoidable} cost becomes ultimately obsolete. However, the success of recent attacks based on ransomware cryptoworms such as WannaCry or NotPetya may indicate that rather than one perfectly executed fatal attack, the dominant attack pattern is actually multiple limited knowledge or even automated zero-knowledge attacks. In this case, again, better average case performance might be more suitable.

% \FloatBarrier
% \subsection{Evaluation of the Learning Process}
% TODO: Some plots episodes versus average reward, 

\FloatBarrier
\section{Conclusion and Future Work}\label{conclusion}
In this paper we formalize the backup strategy against adversaries problem as a Markov decision process model that can be efficiently solved using reinforcement learning techniques. 

By using the $q$-value, a natural discrepancy metric, as basis for our reward function we suggest a learning algorithm based on deep deterministic policy gradients that learns to minimize discrepancy over the run time of the backup scheme.

In experiments with our prototype implementation we demonstrate that our suggested algorithm produces almost identical backup schemes for $k<5$, while obtaining better mean reward and $q$-value scores for higher $k$. 
Due to the universal function approximation property of the policy network, the algorithm is not just a fixed set of update rules and step sizes but can also deal with deviations from the schedule (as they might occur in real world scenarios) and is able to dynamically drive the backup scheme back to the optimal path.

We think there is great potential in the mathematical modelling and formalization of problems in IT Security and have provided one example how reinforcement learning algorithms and consequently (semi-)autonomous agents could help to guide planning and security processes in the future.

%%%%%%%%%%%% Supplementary Methods %%%%%%%%%%%%
%\footnotesize
%\section*{Methods}

%%%%%%%%%%%%% Acknowledgements %%%%%%%%%%%%%
\footnotesize
\section*{Acknowledgements}
We thank Bar-On et al. \cite{bs} for providing us with their implementation and in particular Rani Hod for valuable information on how to use it.
%%%%%%%%%%%%%%   Bibliography   %%%%%%%%%%%%%%
\normalsize
\bibliography{main}

\begin{thebibliography}{16}
\providecommand{\natexlab}[1]{#1}
\providecommand{\url}[1]{\texttt{#1}}
\expandafter\ifx\csname urlstyle\endcsname\relax
  \providecommand{\doi}[1]{doi: #1}\else
  \providecommand{\doi}{doi: \begingroup \urlstyle{rm}\Url}\fi

\bibitem[Upadhyaya and Jain(2016)]{bitcoin_paper}
R.~Upadhyaya and A.~Jain.
\newblock Cyber ethics and cyber crime: A deep dwelved study into legality,
  ransomware, underground web and bitcoin wallet.
\newblock In \emph{2016 International Conference on Computing, Communication
  and Automation (ICCCA)}, pages 143--148, April 2016.

\bibitem[Bringmann et~al.(2013)Bringmann, Doerr, Neumann, and
  Sliacan]{checkpoint_worst_case}
Karl Bringmann, Benjamin Doerr, Adrian Neumann, and Jakub Sliacan.
\newblock Online checkpointing with improved worst-case guarantees.
\newblock volume~27, 02 2013.

\bibitem[Bar-On et~al.(2018)Bar-On, Dinur, Dunkelman, Hod, Keller, Ronen, and
  Shamir]{bs}
Achiya Bar-On, Itai Dinur, Orr Dunkelman, Rani Hod, Nathan Keller, Eyal Ronen,
  and Adi Shamir.
\newblock {Tight Bounds on Online Checkpointing Algorithms}.
\newblock In \emph{45th International Colloquium on Automata, Languages, and
  Programming (ICALP 2018)}, volume 107, pages 13:1--13:13, 2018.

\bibitem[Young(1974)]{young_first_1974}
John~W Young.
\newblock A first order approximation to the optimum checkpoint interval.
\newblock \emph{Communications of the ACM}, 17\penalty0 (9):\penalty0 530--531,
  1974.

\bibitem[Gelenbe(1979)]{gelenbe_optimum_1979}
Erol Gelenbe.
\newblock On the optimum checkpoint interval.
\newblock \emph{Journal of the ACM}, 26\penalty0 (2):\penalty0 259--270, 1979.

\bibitem[Nakamura et~al.(2003)Nakamura, Qian, Fukumoto, and
  Nakagawa]{nakamura_optimal_2003}
S~Nakamura, C~Qian, S~Fukumoto, and T~Nakagawa.
\newblock Optimal backup policy for a database system with incremental and full
  backups.
\newblock \emph{Mathematical and computer modelling}, 38\penalty0
  (11-13):\penalty0 1373--1379, 2003.

\bibitem[Lim et~al.(2011)Lim, Lee, Lee, and Lee]{lim_power-aware_2011}
Sung-Hwa Lim, Se~Won Lee, Byoung-Hoon Lee, and Seongil Lee.
\newblock Power-aware optimal checkpoint intervals for mobile consumer devices.
\newblock \emph{IEEE Transactions on Consumer Electronics}, 57\penalty0
  (4):\penalty0 1637--1645, 2011.

\bibitem[Beck(1987)]{beck1987}
J{\'o}zsef Beck.
\newblock Irregularities of distribution. i.
\newblock \emph{Acta Mathematica}, 159\penalty0 (1):\penalty0 1--49, 1987.

\bibitem[Lillicrap et~al.(2016)Lillicrap, Hunt, Pritzel, Heess, Erez, Tassa,
  Silver, and Wierstra]{ccwdr}
Timothy~P. Lillicrap, Jonathan~J. Hunt, Alexander Pritzel, Nicolas Heess, Tom
  Erez, Yuval Tassa, David Silver, and Daan Wierstra.
\newblock Continuous control with deep reinforcement learning.
\newblock 2016.
\newblock URL \url{http://arxiv.org/abs/1509.02971}.

\bibitem[Mnih et~al.(2013)Mnih, Kavukcuoglu, Silver, Graves, Antonoglou,
  Wierstra, and Riedmiller]{mnih}
Volodymyr Mnih, Koray Kavukcuoglu, David Silver, Alex Graves, Ioannis
  Antonoglou, Daan Wierstra, and Martin~A. Riedmiller.
\newblock Playing atari with deep reinforcement learning.
\newblock \emph{CoRR}, abs/1312.5602, 2013.
\newblock URL \url{http://arxiv.org/abs/1312.5602}.

\bibitem[Sutton et~al.(1999)Sutton, McAllester, Singh, and Mansour]{sutton_ac}
Richard~S. Sutton, David McAllester, Satinder Singh, and Yishay Mansour.
\newblock Policy gradient methods for reinforcement learning with function
  approximation.
\newblock In \emph{Proceedings of the 12th International Conference on Neural
  Information Processing Systems}, NIPS'99, pages 1057--1063, Cambridge, MA,
  USA, 1999. MIT Press.
\newblock URL \url{http://dl.acm.org/citation.cfm?id=3009657.3009806}.

\bibitem[Hausknecht and Stone(2016)]{param_act_space}
Matthew Hausknecht and Peter Stone.
\newblock Deep reinforcement learning in parameterized action space.
\newblock In \emph{Proceedings of the International Conference on Learning
  Representations (ICLR)}, May 2016.

\bibitem[Silver et~al.(2014)Silver, Lever, Heess, Degris, Wierstra, and
  Riedmiller]{dpg_silver}
David Silver, Guy Lever, Nicolas Heess, Thomas Degris, Daan Wierstra, and
  Martin~A. Riedmiller.
\newblock Deterministic policy gradient algorithms.
\newblock In \emph{ICML}, volume~32 of \emph{JMLR Workshop and Conference
  Proceedings}, pages 387--395. JMLR.org, 2014.
\newblock URL
  \url{http://dblp.uni-trier.de/db/conf/icml/icml2014.html#SilverLHDWR14}.

\bibitem[Schaul et~al.(2016)Schaul, Quan, Antonoglou, and
  Silver]{schaul2015prioritized}
Tom Schaul, John Quan, Ioannis Antonoglou, and David Silver.
\newblock Prioritized experience replay.
\newblock In \emph{4th International Conference on Learning Representations,
  {ICLR} 2016, San Juan, Puerto Rico, May 2-4, 2016, Conference Track
  Proceedings}, 2016.
\newblock URL \url{http://arxiv.org/abs/1511.05952}.

\bibitem[Browne et~al.(2012)Browne, Powley, Whitehouse, Lucas, Cowling,
  Rohlfshagen, Tavener, Perez, Samothrakis, and Colton]{browne_survey_2012}
Cameron~B Browne, Edward Powley, Daniel Whitehouse, Simon~M Lucas, Peter~I
  Cowling, Philipp Rohlfshagen, Stephen Tavener, Diego Perez, Spyridon
  Samothrakis, and Simon Colton.
\newblock A survey of monte carlo tree search methods.
\newblock \emph{IEEE Transactions on Computational Intelligence and AI in
  games}, 4\penalty0 (1):\penalty0 1--43, 2012.

\bibitem[Chollet et~al.(2015)]{keras}
Fran\c{c}ois Chollet et~al.
\newblock Keras.
\newblock \url{https://keras.io}, 2015.

\end{thebibliography}

%%%%%%%%%%%%  Supplementary Figures  %%%%%%%%%%%%
\clearpage
\section{Experiments Details}\label{appendix:experiments}
The implementation is written in Keras \cite{keras} and follows the actor, critic architecture. For training of both actor and critic the Adam optimizer was used. Any additional hyperparameters are collected in the  Table \ref{table:hyper_params}.
\begin{table}
    \begin{center}
    \begin{tabular}{lll}
        \toprule
        \textbf{Parameter}    &  \textbf{Description}  & \textbf{Value} \\
        \midrule
        $N_{\text{eps}}$ & number of episodes & 600 \\
        $N_{\text{steps}}$ & number of steps/actions per episode & 10000 \\
        $\lambda$ & reward scale & 5.0 \\
        $l_r$ & learning rate critic & 0.001 \\
        $l_a$ & learning rate actor & 0.0001 \\
        $\tau$ & lag parameter target update & 0.001 \\
        $\gamma$ & reward discount rate & 0.99 \\
        $\epsilon_{d,\text{min}}$ & minimum epsilon for \\
        &$\epsilon$-greedy exploration (discrete) & 0.1\\
        $\Delta\epsilon_d$ & epsilon increments for linear annealing& 1/(3e6) \\
        $\epsilon_{c,\text{min}}$ & minimum epsilon for  \\
        & continuous exploration (noise scale) & 0.1\\
        $\Delta\epsilon_c$ & epsilon increments for linear annealing& 1/(3e6) \\
        $\mu_{OU}$ & Ornstein-Uhlenbeck mean & 0.0 \\
        $\theta_{OU}$ & Ornstein-Uhlenbeck mean reversion speed & 0.3 \\
        $\sigma_{OU}$ & Ornstein-Uhlenbeck standard deviation & 0.1 \\
        $N_{B}$ & size of the replay buffer &1000000 \\
        $N_{b}$ & Minibatch Size &32 \\
        \bottomrule
    \end{tabular}
    \end{center}
    \caption{Hyperparameters\label{table:hyper_params}}
\end{table}
%%%%%%%%%%%%%%%%   End   %%%%%%%%%%%%%%%%
%\end{multicols}  % Method B for two-column formatting (doesn't play well with line numbers), comment out if using method A
\end{document}